# Professional Development of Teachers Using Cloud Services During Non-formal Education

Svitlana Lytvynova, Oksana Melnyk

Institute of Information Technologies and Learning Tools of NAES of Ukraine,
9 Maksyma Berlyns'koho St, 04060 Kyiv, Ukraine

s_litvinova@i.ua, ok_melnyk@ukr.net

**Abstract.** The rapid development of cloud services and their implementation in secondary education require an increase in the IC-competence of teachers during non-formal education. The implementation of cloud services will make it possible to create some conditions for learning mobility of all participants of teaching and learning activities. This article deals with the organization of teachers' training during non-formal learning. The paper analyzes the main forms of the organization of teachers' learning (workshops, trainings and summer schools). The special features of the non-formal teachers' ICT training include the availability of high-speed Internet and some computer equipment. The obtained basic and additional services allow teachers to make the extensive use of cloud services for different activities, namely the organization of students' group work and "inverted learning"; team-work on projects, assistance during homework; preparation for contests; conducting web-quests.

**Key Terms.** TeachingMethodology, TeachingProcess, KnowledgeManagementMethodology, QualityAssuranceMethodology, StandardizationProcess

## 1   Introduction

The draft of The Concept of Education in Ukraine for 2015-2025 noted that in the period of independency it was accumulated a lot of systematic problems in the Ukrainian education sector. They include the following: reducing of the logistical support of educational institutions; teachers' aging; lack of young specialists; inefficient, over-centralized, out-of-date management and financing of the system; growing inequality in access to education; decline of students' knowledge and skills quality; obsolescence of technology and teaching methods; lack of effective system for education quality control; reduction in textbooks' quality and lack of advanced technolo-gies. In addition, there are unresolved issues: the unsatisfactory state of computer equipment; the poor quality and accessibility of the Internet, the lack of systematic updating of ICT applications in the educational process.

In the current information age, the transition from education when a teacher is a central figure for information transmission to person-oriented education is one of the most significant educational change [6].

Most of the mentioned problems can be solved by the introduction of cloud services in education, which will provide educational mobility of educational process participants and the virtualization of organizational and methodological components of educational process [8, p.209].

A detailed analysis of the foreign projects in Russia, Germany, Czech Republic, Australia, China, Israel, Africa, Singapore, Brazil, Colombia, Azerbaijan and the United States showed that the cloud-oriented learning environments are used by the teachers and students of foreign countries to improve the educational process, the access to training materials, schedules of work, curriculums for the revitalization of students' activities, to make easily the educational process during a quarantine period, the process of receiving homework and distance learning [8, p. 27].

Thus, cloud-oriented learning environments have a number of advantages for educational institutions in the organization of educational process and the use of learning technologies [9, p. 13].

Therefore, a teachers' training is crucial for an application of cloud technologies. It can be done through non-formal education.

The educational trends in the global space are characterized by considerable support and development of non-formal education. The term "non-formal education" is widely used in the scientific world community today. The understanding of "non-formal education" at the European level is gaining its significance for Ukraine during the current period of integration into the EU.

Zinchenko S. offers the following characteristics of non-formal education: social character of leadership; variability of educational programs and terms; combination of scientific and applied knowledge; voluntary of education; systematic character of education; goal-oriented activity of learners; orientation to satisfaction the educational needs of specific social and professional groups; creating comfortable learning environment for adults' communication; possibility of psychological protection during some social changes [15].

According to the research of N. Terjokhina, non-formal education in Ukraine covers the following spheres: non-school education; postgraduate education and adult education; school and student self-government (through the acquiring of organizational, communication, management skills); educational initiatives to develop additional skills and abilities (language and computer courses, hobby clubs, etc.); civic education (various activities of non-governmental organizations); folk high schools [14].

Summarizing the organizational structure of adult education in current Ukraine, Sighajeva L. identifies three forms of non-formal education: courses; hobby clubs; non-governmental organizations [13].

We consider that it is very important for the further development of non-formal education in Ukraine to take into account the recommendations for the development of non-formal education for the EU members and countries, which are going to be its

members, proposed by the Committee of Ministers of the Council of Europe. They include:

1. The development of effective standards, which recognize non-formal education as an integral part of general education, which are quality criteria of the educational process during non- formal education.

2. Providing information and examples of best practices, teaching methods, lists of skills, experience and knowledge, obtained through non-formal education should be developed.

## 2      Research Methods

Analysis, theoretical systematization of data; synthesis and evaluation of experimental work; the factor-criteria model to determine the level of teachers' IC-competence on the use of cloud services and the experiment, which was held among 190 subject teachers of Ukrainian secondary schools.

It was determined the ranking coefficient for each criterion to calculate the teachers' IC-competence level in the use of cloud services (Table 1).

**Table 1.** Factors and Ranking Coefficients

| Factors | Ranking coefficients |
|---|---|
| Understanding of the role and use of cloud services in education | 0.158 |
| Basic knowledge about cloud services | 0.16 |
| Use of cloud services in professional activities | 0.17 |
| Ability for collaboration and self-education | 0.174 |
| Use of basic cloud services | 0.172 |
| Use of different forms of learning activity | 0.178 |

## 3      Research Findings

There is a widespread adoption of cloud services in general secondary education. They are the main tools for the efficient organization of cooperation and cooperative work between students and teachers now. That is why the organization of teachers' training during non-formal education is necessary to increase their level of competence in the sphere of cloud services' application.

We can define the following forms of non-formal education: workshops, trainings, summer schools.

Teachers should know theoretical basics and the capabilities of the services during the training seminars, where they will be offered short practical work on the use of several services for the improvement of educational process or the organization of cooperation with students and to drill the material.

The purpose of the training is to improve the skills of working with certain services, so it gives attention on some problematic issues and extending teachers' knowledge.

During the summer schools teachers are given an opportunity not only to combine the two forms of competence development mentioned above, but also unlimited communication with colleagues, team-work on projects, familiarization with the latest innovations of IT industry and pedagogy.

The high-speed Internet and a gadget (laptop, tablet, etc.) are the two components that can be attributed to the special features of the organization of the mentioned training.

The training can be carried out on the basis of services in the cloud-oriented learning environment designed for teachers' training or in schools.

Cloud services include programs that are provided on demand in on-line mode, such as basic and additional ones. The basic programs include calendar, corporate email and document storage, basic office software (Word, Excel, PowerPoint and Forms). The additional cloud services are OneNote, Sway, GeoGebra, OneNote Classroom, the corporate electronic network Yammer, Power BI, Delve, Video, Project etc.

The pedagogical experiment was conducted in secondary schools of Ukraine, working within the framework of the experimental work on topic "Cloud services in education" (order of the Ministry of Education and Science of Ukraine of 21.05.2014 №629, scientific guide Lytvynova S.), namely in 18 experimental schools (Kyiv, Lugansk, Khmelnytsky, Dnipropetrovsk, Zhytomyr, Sumy, Vinnytsia, Ternopil regions). To collect data and determine the levels of teachers' IC-competence, it was developed the portal "The Research System" (http://expert.obolon365.net). 190 subject teachers took place in the experiment.

The teacher training was conducted during non-formal education by three stages.

The first stage "Designing of cloud-oriented learning environment" was conducted in the form of a training workshop (8 hours), during which each teacher designed his "cloud space".

On the second stage the teachers were offered intensive implementation of services during the summer school (40 hours). An educational project was developed in the result of the work.

At the third stage the teachers presented their best practices of the use of cloud services for educational purposes during a workshop.

Let us examine the advantages, specific features and disadvantages of the teachers' trainings within non-formal education on the use of cloud services (Table 2).

As a result of the experiment, it was found that the volume of use of cloud services by the teachers has almost doubled, and the level of IC-competence respectively: teachers' understanding of the cloud services increased by 41%, basic knowledge about cloud services - by 45%, the use of cloud services in professional work – by 46%, ability to cooperate with the help of cloud services – by 43%, the use of basic services in students' education – by 45%, the use of various forms of training activities with the help of cloud services - by 44% (Fig.1).

**Table 2.** Advantages, Features and Disadvantages

| Type of non-formal education | Advantages | Features | Disadvantages |
|---|---|---|---|
| Workshops | Demonstration of practical application of services in the real-world | Duration up to 4 hours | An additional training of school teachers to present the use of services is required |
| Trainings | Presentation or training on the use of new services for educational purposes | Duration 4-6 hours | No more than 3-4 questions are considered, an additional training of trainers is required |
| Summer schools | Improving the knowledge on the use of certain services, consulting, communication with experts | Duration 40 hours | Public events with 60-120 participants; a special accommodation for training teachers, a broadband Internet connection and an additional training of trainers are required |

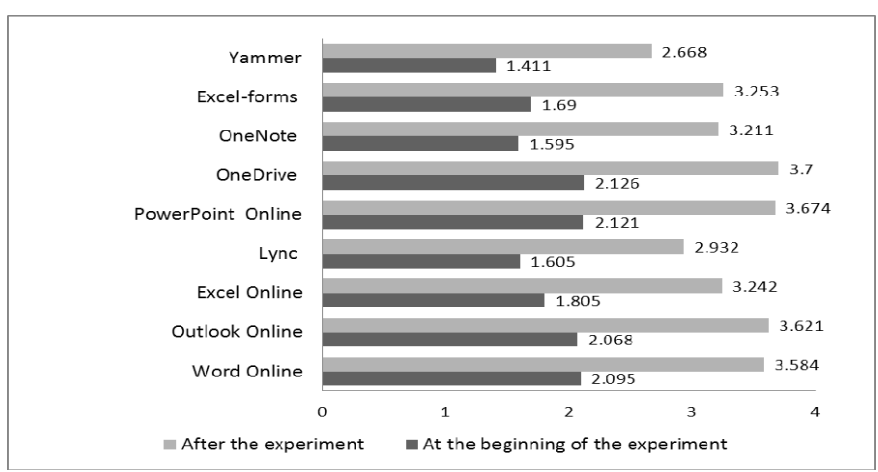

**Fig. 1.** The volume of using basic cloud services by the teachers

The teachers' training of new information and communication technologies, in particular, cloud services during non-formal education, made it possible to summarize the following: teachers actively improve their ICT qualification; there is no problem of a lack of gadgets; the systematic training in three forms gives significant results in the development of teachers' IC-competence; cloud services become the effective tools in the organization of cooperation and cooperative work between teachers and pupils.

## 4      Discussion

In the process of the selection of cloud services efficiency criteria, the parameters for an assessment of teachers' educational activity provoked a discussion. Some experts have focused their attention on the fact that cloud services should be used for the coordination of training projects and students' self- preparation for lessons. However, most experts emphasized the need for improving the educational process and defined such criteria as the organization of personalized learning, working in small groups, on-line communication with students.

Teachers' broad practical experience with gifted students became a barrier for the use of cloud services during the students' preparation for school contests (Olympiads) and coordination of scientific works of Minor Academy of Sciences (MAS) members. The results of the teachers' survey are shown in Fig. 2.

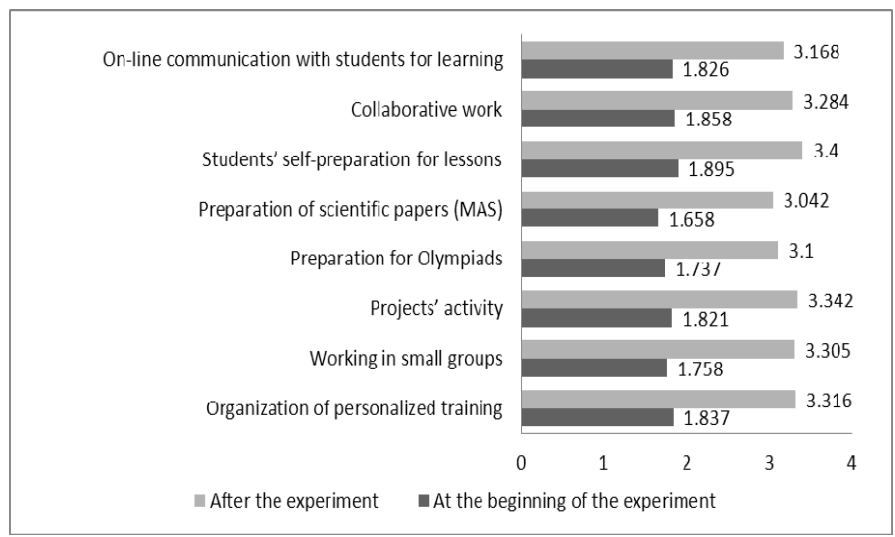

**Fig. 2.** The results of the teachers' survey

Analyzing the obtained results, it can be stated that teachers' lifelong learning is becoming a necessity. The pupils of the XXI century require new technologies in the organization of teaching process, due to the requirements to the quality of secondary education. The emergence of new technologies, such as cloud services, gave an impulse for an active development of teachers' non-formal education.

## 5 Conclusions and Prospects of Further Research

Cloud services contribute to the improvement of the educational process. The development of these services can be arranged through non-formal education of three forms: practical workshops, trainings, summer schools.

The implementing of person-oriented education is a prerequisite for learning with the help of cloud services. There is no need for the absolute traditional teaching methods. The situations, when a student adapt to a teacher, could be avoided [8].

During the active use of cloud services the level of teachers' IC-competence increased from the beginner to average (Table 3).

The efficiency of enhancing teachers' IC-competence in the sphere of learning new technologies is 40% and more. This allows the teachers to make extensive use of cloud services for different activities, namely the organization of students group work, team-work on projects, "inverted learning"; assistance during homework; preparation for contests; conducting web-quests.

**Table 3**. Criteria and levels of teachers' IC- competence

| Criteria | At the beginning of the experiment | After the experiment |
|---|---|---|
| Technological literacy | beginner level | average level |
| Extending of knowledge | beginner level | average level |
| ICT-creativity | beginner level | average level |

So, the organizing of teachers' training on the use of cloud services in order to improve their ICT-competence in collaboration, global communication and cooperative work is a new direction in teachers' training.

On the other hand, the mechanism of teachers' training within non-formal education on the use of cloud services is not perfect yet. Therefore, the further research of this issue is required.